\documentclass[preprint,pra,onecolumn]{revtex4}%
\usepackage{amsfonts}
\usepackage{amsmath}
\usepackage{amssymb}
\usepackage{graphicx}
\usepackage{caption2}%
\providecommand{\U}[1]{\protect\rule{.1in}{.1in}}
\renewcommand\caption{}

\begin{document}
\title{Practical decoy state method in quantum key distribution with heralded single
photon source}
\author{Qin Wang$^{1}$}
\email{wq0305@ustc.edu}
\author{Xiang-Bin Wang$^{2}$}
\author{Guang-Can Guo$^{1}$}
\affiliation{$^{1}$Key Laboratory of Quantum Information, Department of Physics, University
of Science and Technology of China, Hefei 230026, People's Republic of China}
\affiliation{$^{2}$Department of Physics, Tsinghua University, Beijing 100084, China}

\begin{abstract}
We propose a practical decoy state method with heralded single photon source
for quantum key distribution (QKD). In the protocol, 3 intensities are used
and one can estimate the fraction of single-photon counts. The final key rate
over transmission distance is simulated under various parameter sets. Due to
the lower dark count than that of a coherent state, it is shown that a
3-intensity decoy-state QKD with a heralded source can work for a longer
distance than that of a coherent state.

PACS number(s): 03.67.Dd, 42.65.Yj, 03.67.Hk

\end{abstract}
\maketitle

\section{Introduction}

As is well known that, a two-mode light source, such as the state from a
parametric down conversion (PDC) or a two-mode squeezed state takes a very
important role in quantum information processing (QIP)\cite{klys}. Since the
two modes always have same number of photons, one mode can be used to indicate
the state of the other, therefore it is also called heralded single photon
source (HSPS). Moreover, recent years, due to its wide applications in many
fields, such as in quantum information and quantum radiometry \textit{etc.,
}the technology on how to efficiently obtain HSPS has been developed to a high
level \cite{fase,albo,kurt,cast,ljun,bovi,pitt,cast2,cast3}.

In the past few years, quantum key distribution has attracted extensive
attentions for its unconditional security compared with conversional
cryptography \cite{benn,maye,shor,eker,deut,deut2}. However, there still exist
some limitations in practice, such as imperfect single source, large loss
channel and inefficient detectors \textit{etc}.. Under such limitations, one
serious threaten to the security is the so called photon number splitting
attack\cite{PNS1,PNS2,lutk}. Fortunately, a number of methods and proposals
have been presented for secure QKD even with these imperfections. These
include the mixed protocol\cite{scran}, the strong-reference light
method\cite{kko}, and the decoy-state method\cite{gott,hwan,wang,lo}. And in
this paper, what we are interested with is the decoy-state method. Firstly,
according to the separate result of ILM-GLLP, we can distill a secure final
key even an imperfect source is used, provided that we know the lower bound of
fraction of single photon-counts\cite{gott}. The non-trivial problem on how to
verify a tight lower bound was not answered then. Later, Hwang\cite{hwan}
proposed the decoy state method to verify such a lower faithfully.

After that, decoy state method has been advanced by several researchers
\cite{wang,lo,wang2,ma1,harr}. The main idea there is to randomly change the
intensity of each pulses among different values, and then deduce the lower
bound of fraction of single-photon counts according to the observed counts of
different intensities. In particular, it has been shown that, one can make a
very tight estimation, i.e., the estimated lower bound is only a bit larger
than the true value therefore a good final key rate can be obtained by only
using 3 intensities, $0,\mu,\mu^{\prime}$\cite{wang} or 4
intensities\cite{wang2}.

However, using the coherent states, the dark count will be significant given a
distance longer than 100 kilometers. Naturally, one may think about using HSPS
to decrease the effect of dark count. In fact, as it has been shown recently,
an HSPS can indeed raise the distance for QKD if one knows the channel
transmittances of each photon number states exactly\cite{hori}. However,
knowing these exactly requires using infinite number of intensities. This
seems to be an impossible task in practice. Very recently, it is proposed to
use photon-number-resolving detectors to do decoy-state QKD with
HSPS\cite{gang}.

In this paper, we will propose a practical decoy state method with HSPS. We
only need 3 intensities and normal yes-no single-photon detectors at Bob's
side. That is to say, we only assume the technologies that have already
adopted in the existing set-ups.

This paper is organized as follows: In Sec. II, starting from the two-mode
state, we shall then present the main result of our protocol, i.e., the lower
bound of single-photon counts. In Sec. III, we estimate the QKD distance of
our method in various settings. This work is concluded in Sec IV.

\section{Heralded Single photon source}

Given a two-mode state of the form
\begin{equation}
|\chi\rangle=\mathrm{cosh}^{-1}\chi\sum_{n=0}^{\infty} e^{in\theta}
\mathrm{tanh}^{n}\chi|n,n\rangle.
\end{equation}
%Since $\theta$ is determined by the phase of pump light which is
%totally random, the state is actually a mixed state of
%\begin{equation}
%\rho_\chi= \int_0^{2\pi}|\chi\rangle\langle\chi| {\rm d}\theta
%=\frac{1}{{\rm cosh}^2\chi}\sum {\rm tanh}^{2n}\chi
%|n,n\rangle\langle n,n|.
%\end{equation}
The averaged photon number in one mode is $\mathrm{sinh}^{2}\chi$ and we shall
use this value to indicate the pulse intensity, i.e., when we say that we use
intensities of $0,\mu,\mu^{\prime}$ for the two mode state as described by
Eq.(1), we mean that $\mathrm{sinh}^{2}\chi=0,\mu,\mu^{\prime}$, respectively.

As indicated in Ref. \cite{lutk}, after triggering out one of a photon pair,
the other mode is basically a thermal field of distribution :
\begin{equation}
\rho_{x}=\frac{1}{P_{post}(x)}\left\{  \frac{d_{A}}{1+x}|0\rangle
\langle0|+\sum_{n=1}^{\infty}\left[  1-\left(  1-\eta_{A}\right)  ^{n}\right]
\frac{x^{n}}{\left(  1+x\right)  ^{n+1}}|n\rangle\langle n|\right\}  ,
\end{equation}
where $x$ is the mean photon number of one mode (before triggering), $\eta
_{A},d_{A}$ for the detection efficiency and dark count rate of Alice's
detector and the post-selection probability is $P_{post}(x)=\frac{d_{A}}%
{1+x}+\frac{x\eta_{A}}{1+x\eta_{A}}$. The detectors assumed in our protocol
here are threshold detectors, i.e., the outcome of each individual measurement
is either clicking or not clicking.

In the protocol, we request Alice to randomly change the intensities of her
pump light among 3 values, so that the intensity of one mode of the two mode
source is randomly changed among $0,\mu,\mu^{\prime}$ (and $\mu^{\prime}>\mu
$). We define $Y_{n}$ to be the yield of a n-photon state, i.e., the
probability that Bob's detector click whenever Alice sends out state
$|n\rangle$. We also denote the yield state $\rho_{\mu},\rho_{\mu^{\prime}}$
by $Y_{\mu},Y_{\mu^{\prime}}$. In the protocol, one can immediately know the
value of $Y_{0}$ by watching the counts of vacuum pulses, and one can also
know the value of $Y_{\mu},Y_{\mu^{\prime}}$ by watching the counts caused by
pulses of intensity $\mu,\mu^{\prime}$ respectively. In particular, suppose
there are $N_{x}$ pulses for state $\rho_{x}$ and $N_{xt}$ of them are
triggered. During the time windows of these $N_{xt}$ pulses, Bob has observed
$n_{x}$ clicks. Then we have $Y_{x}=n_{x}/N_{xt}$ according to our definition
of yield. We want to deduce the value of $Y_{1}$ based on the known parameters
$Y_{0}$, $Y_{\mu}$ and $Y_{\mu}^{\prime}$. Similar to the case of coherent
states\cite{wang}, we can verify the lower bound of $Y_{1}$ by the following
constraints:
\begin{equation}
\tilde{Y}_{\mu}=Y_{0}\frac{d_{A}}{1+\mu}+%
%TCIMACRO{\dsum _{i=1}^{\infty}}%
%BeginExpansion
{\displaystyle\sum_{i=1}^{\infty}}
%EndExpansion
Y_{n}\left[  1-\left(  1-\eta_{A}\right)  ^{n}\right]  \frac{\mu^{n}}{\left(
1+\mu\right)  ^{n+1}},
\end{equation}
and
\begin{equation}
\tilde{Y}_{\mu^{\prime}}=Y_{0}\frac{d_{A}}{1+\mu^{\prime}}+%
%TCIMACRO{\dsum _{i=1}^{\infty}}%
%BeginExpansion
{\displaystyle\sum_{i=1}^{\infty}}
%EndExpansion
Y_{n}\left[  1-\left(  1-\eta_{A}\right)  ^{n}\right]  \frac{{\mu^{\prime}%
}^{n}}{\left(  1+\mu^{\prime}\right)  ^{n+1}},
\end{equation}
and $\tilde{Y}_{x}=(N_{xt}/N_{x})Y_{x},$ $x=\mu,\mu^{\prime}$. These two
equations lead to%
\begin{align}
&  (1+\mu)\left(  \frac{\mu^{\prime}}{1+\mu^{\prime}}\right)  ^{2}\tilde
{Y}_{\mu}-(1+\mu^{\prime})\left(  \frac{\mu}{1+\mu}\right)  ^{2}\tilde{Y}%
_{\mu^{\prime}}\nonumber\\
&  =Y_{0}\frac{d_{A}}{1+\mu}\left(  \frac{\mu^{\prime}}{1+\mu^{\prime}%
}\right)  ^{2}-Y_{0}\frac{d_{A}}{1+\mu^{\prime}}\left(  \frac{\mu}{1+\mu
}\right)  ^{2}\nonumber\\
&  +\eta_{A}Y_{1}\left[  \frac{\mu}{1+\mu}\left(  \frac{\mu^{\prime}}%
{1+\mu^{\prime}}\right)  ^{2}-\frac{\mu^{\prime}}{1+\mu^{\prime}}\left(
\frac{\mu}{1+\mu}\right)  ^{2}\right] \nonumber\\
&  +\sum_{n=3}^{\infty}Y_{n}\left[  1-(1-\eta_{A})^{n}\right]  \left[
\frac{\mu^{n}{\mu^{\prime}}^{2}}{(1+\mu)^{n}(1+\mu^{\prime2})}-\frac
{{\mu^{\prime}}^{n}{\mu}^{2}}{(1+\mu^{\prime n}(1+\mu^{2})}\right]  .
\label{media}%
\end{align}

It is easy to see that for any $n\geq3$, $\frac{\mu^{n}{\mu^{\prime}}^{2}%
}{(1+\mu)^{n}(1+\mu^{\prime2})}-\frac{{\mu^{\prime}}^{n}{\mu}^{2}}%
{(1+\mu^{\prime n}(1+\mu^{2})}<0$ given that $\mu^{\prime}>\mu$. Therefore
Eq.(\ref{media}) leads to the following inequality:
\begin{align}
Y_{1}  &  \geq\left\{  \frac{\mu^{\prime}}{\mu}(1+\mu)^{3}\tilde{Y}_{\mu
}-\frac{\mu}{\mu^{\prime}}(1+\mu^{\prime3}\tilde{Y}_{\mu^{\prime}}-Y_{0}%
d_{A}\left[  \frac{\mu^{\prime}}{\mu}(1+\mu)^{2}-\frac{\mu}{\mu^{\prime}%
}(1+\mu^{\prime2}\right]  \right\}  \times\nonumber\\
&  \lbrack\eta_{A}(\mu^{\prime}-\mu)]^{-1} \label{main1}%
\end{align}
This gives rise to the fraction of single-photon counts for the triggered
pulses of different intensities by the following formula
\begin{equation}
\Delta_{1}(x)=\frac{Y_{1}\eta_{A}x}{Y_{x}P_{post}(x)(1+x)^{2}} \label{main2}%
\end{equation}
and $x$ can be $\mu$ or $\mu^{\prime}$ here. Also, if we have observed the
quantum bit-flip rate (QBER) for triggered pulses of intensity $x$ is $E_{x}$,
we can upper bound the QBER value for those single-photon pulses by
\begin{equation}
e_{1}\leq\frac{(1+x)^{2}E_{x}\tilde{Y}_{x}-(1+x)Y_{0}d_{A}/2}{Y_{1}\eta_{A}x}.
\end{equation}
Normally, we use the value from $x=\mu$ for a tight estimation of $e_{1}$.
Given all these, we can use the following formula to calculate the final
key-rate of triggered signal pulses:
\begin{equation}
R\geq\frac{Y_{\mu^{\prime}}P_{post}(\mu^{\prime})}{2}\left\{  -f\left(
E_{\mu^{\prime}}\right)  H_{2}\left(  E_{\mu^{\prime}}\right)  +\Delta_{1}%
(\mu^{\prime})\left[  1-H_{2}\left(  e_{1}\right)  \right]  \right\}
\label{krate}%
\end{equation}
where the factor $\frac{1}{2}$ comes from the cost of basis miss-match in
Bennett-Brassard 1984 (BB84) protocol; $f(E_{\mu^{\prime}})$ is a factor for
the cost of error correction given existing error correction systems in
practice. We assume $f=1.2$ here. $H_{2}\left(  x\right)  $ is the binary
Shannon information function, given by
\[
H_{2}\left(  x\right)  =-x\log_{2}(x)-(1-x)\log_{2}(1-x).
\]

\section{Numerical simulation}

In an experiment, we only need to observe the values of $Y_{0},Y_{\mu}%
,Y_{\mu^{\prime}}$ and $E(\mu),E_{\mu^{\prime}}$ and then deduce the lower
bound of fraction of single-photon counts and upper bound QBER of
single-photon pulses by the theoretical results and then one can distill the
secure final key.

Here our goal is to theoretically estimate the final key rate of our protocol
if we really \emph{did} the experiment. In principle, whatever possible
results can be observed in an experiment. In evaluating our the efficiency of
our protocol theoretically, we shall only consider the normal case where there
is no Eve and we calculate the key rate with respect to distance. In order to
make a faithful evaluation, we first need a model to forecast what values for
$Y_{0},Y_{\mu},Y_{\mu^{\prime}},Y_{\mu}$ and $E(\mu),E_{\mu^{\prime}}$
\emph{would} be observed if we \emph{did} the experiment in the case there is
no Eve. After these values are estimated, we can calculate the final key rate
by Eq.(\ref{krate}) therefore the efficiency is evaluated theoretically.

Suppose $\eta$ is the overall transmittance and detection efficiency between
Alice and Bob; $t_{AB}$ is the transmittance between Alice and Bob,
$t_{AB}=10^{-\alpha L/10}$; $\eta_{B}$ is the transmittance in Bob's side,
$\eta=t_{AB}.\eta_{B}$. Therefore normally, the observed value for $Y_{\mu
},Y_{\mu^{\prime}}$ should be around
\begin{equation}
Y_{x}=\frac{1}{P_{post}(x)} \left\{  \frac{d_{A}d_{B}}{1+x}+\sum_{n=1}%
^{\infty}\frac{[1-(1-\eta_{A})^{n}]x^{n}}{(1+x)^{n+1}} [d_{B}+1-(1-\eta
)^{n}]\right\}
\end{equation}
and $x$ can be any of $\mu,\mu^{\prime}$, $d_{B}$ is the dark count rate of
Bob's detector. This leads to
\begin{equation}
\tilde Y_{x}=\frac{d_{A}d_{B}}{1+x}+\sum_{n=1}^{\infty}\frac{[1-(1-\eta
_{A})^{n}]x^{n}}{(1+x)^{n+1}}[d_{B}+1-(1-\eta)^{n}]
\end{equation}
and $\tilde Y_{x}=Y_{x}\cdot P_{post}(x)$ is directly used in Eq.(\ref{main1}%
), Eq.(\ref{main2}) for calculating the single-photon counts.

We use the following for the error rate of an \emph{n-photon} state:
\begin{equation}
e_{n}=\frac{e_{0}d_{B}+e_{d}[1-(1-\eta)^{n}]}{d_{B}+1-(1-\eta)^{n}}%
\end{equation}
where $e_{0}=1/2$, $d_{B}$ is dark count rate of Bob's detectors, $e_{d}$ is
the probability that the survived photon hits a wrong detector, which is
independent of the transmission distance. Below we shall assume $e_{d}$ to be
a constant. Therefore, the observed $E_{\mu}$ value should be around
\begin{equation}
E_{x}=\frac{e_{0}d_{B}}{Y_{x}}+\frac{1}{Y_{x}P_{post}(x)}\sum_{n=1}^{\infty
}\frac{e_{d}x^{n}[1-(1-\eta_{A})^{n}][1-(1-\eta)^{n}]}{(1+x)^{n+1}}.
\end{equation}
In practical implementation of QKD, we often use the non-degenerated
down-conversion to produce photon pairs, with one photon at the wavelength
convenient for detection acting as heralding signal, and the other falls into
the telecommunication windows for optimal propagation along the fiber or in
open air acting as heralded signal. To illustrate the calculation, we assume
the heralding photon is adapted to be 800nm, and the heralded one to be
1550nm. We shall use these formulas and experimental parameters of GYS as
listed in table I to simulate the observed values $Y_{\mu},~Y_{\mu^{\prime}}$
for Eq.(\ref{main1},\ref{main2}) and the calculate the final key rate by
Eq.(\ref{krate}).

We can now calculate the final key generation rate with the assumed observed
values above. For convenience of comparing with the result of coherent states,
we use the same parameters as in GYS \cite{gobb}, shown in Table I, and
$d_{A}=10^{-5}$. \begin{figure}[ptb]
\caption{Table1. Experimental parameters in GYS. }
\begin{center}
\includegraphics[scale=0.80]{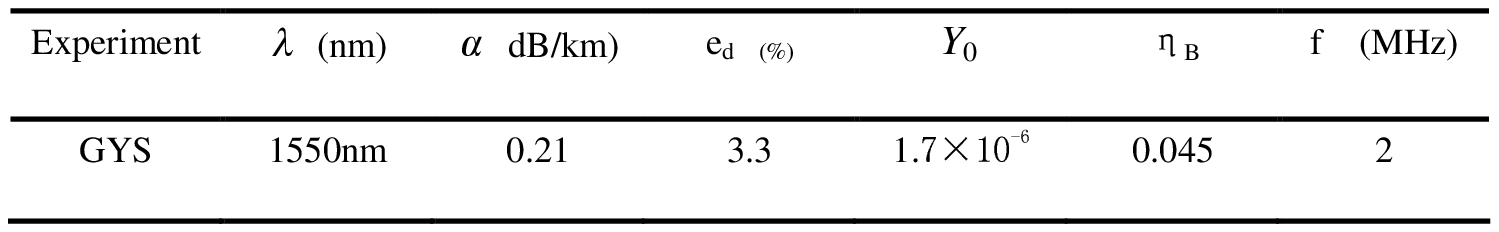}
\end{center}
\end{figure}Our simulation results are shown in Fig. 1, Fig. 2 and Fig. 3.
\begin{figure}[ptb]
\begin{center}
\includegraphics[scale=0.70]{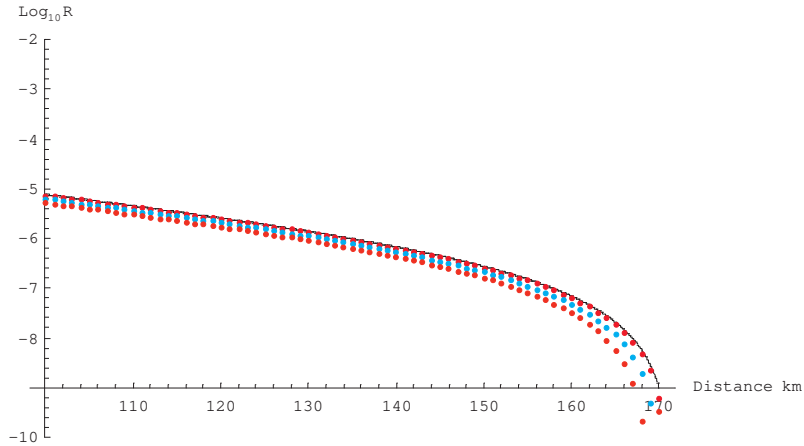}
\end{center}
\caption{Fig1. Final key rates vs transmission distance for decoy state method
with an HSPS source. The black solid line is the ideal result where the
fraction of single-photon counts and QBER of single-photon pulses are known
exactly. The dotted lines are the results of our 3-intensity decoy state
method with $\mu=0.01,0.05,0.10,$ from upper to down. ($\mu^{\prime}$ has the
optimal value at each point.)}%
\label{Fig1}%
\end{figure}

Fig. 1 shows with HSPS the key generation rate against
transmission distance in the asymptotic decoy state method and in
our practical two decoy state method, we use different intensity
of weak decoy state ($\mu=0.01,0.05,0.10$) respectively. From the
curves in Fig. 1, we can see that our 3-intensity decoy state
method can asymptotically approach the theoretical limit of ideal
case.

\begin{figure}[ptb]
\begin{center}
\includegraphics[scale=0.50]{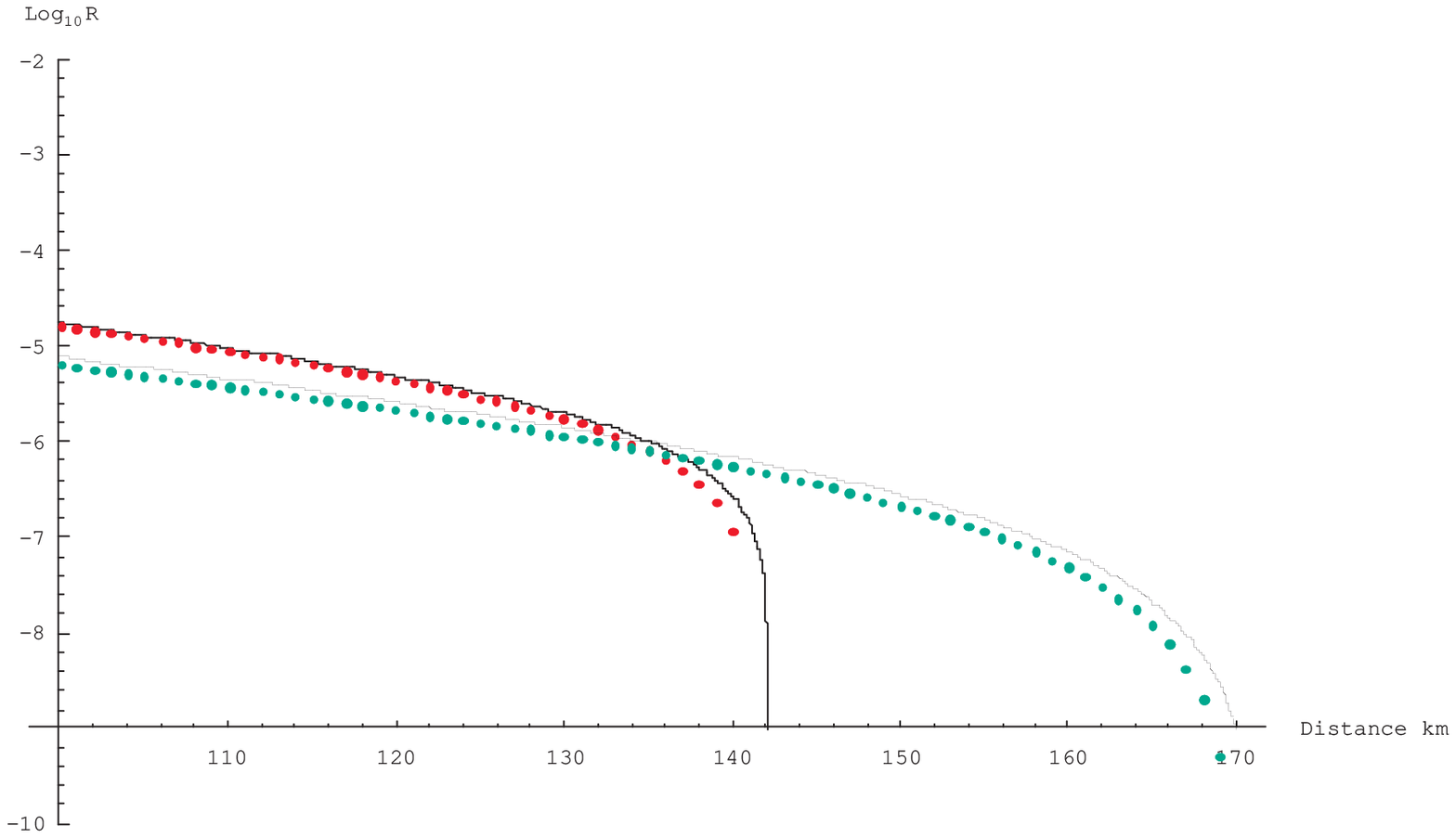}
\end{center}
\caption{Fig2. Comparison of final key rate between a 3-intensity decoy-state
protocol with an HSPS source and the one with a coherent state source. The
solid lines are for the ideal results and the dotted lines are for 3-intensity
protocols. The green dotted line represents the result of an HSPS source (with
$\eta_{A}=0.8$ , $\mu=0.05$), and the red dotted line represents the results
of a coherent state source (with $\mu=0.05$). ($\mu^{\prime}$ has the optimal
value at each point.)}%
\label{Fig2}%
\end{figure}

\begin{figure}[ptb]
\begin{center}
\includegraphics[scale=0.50]{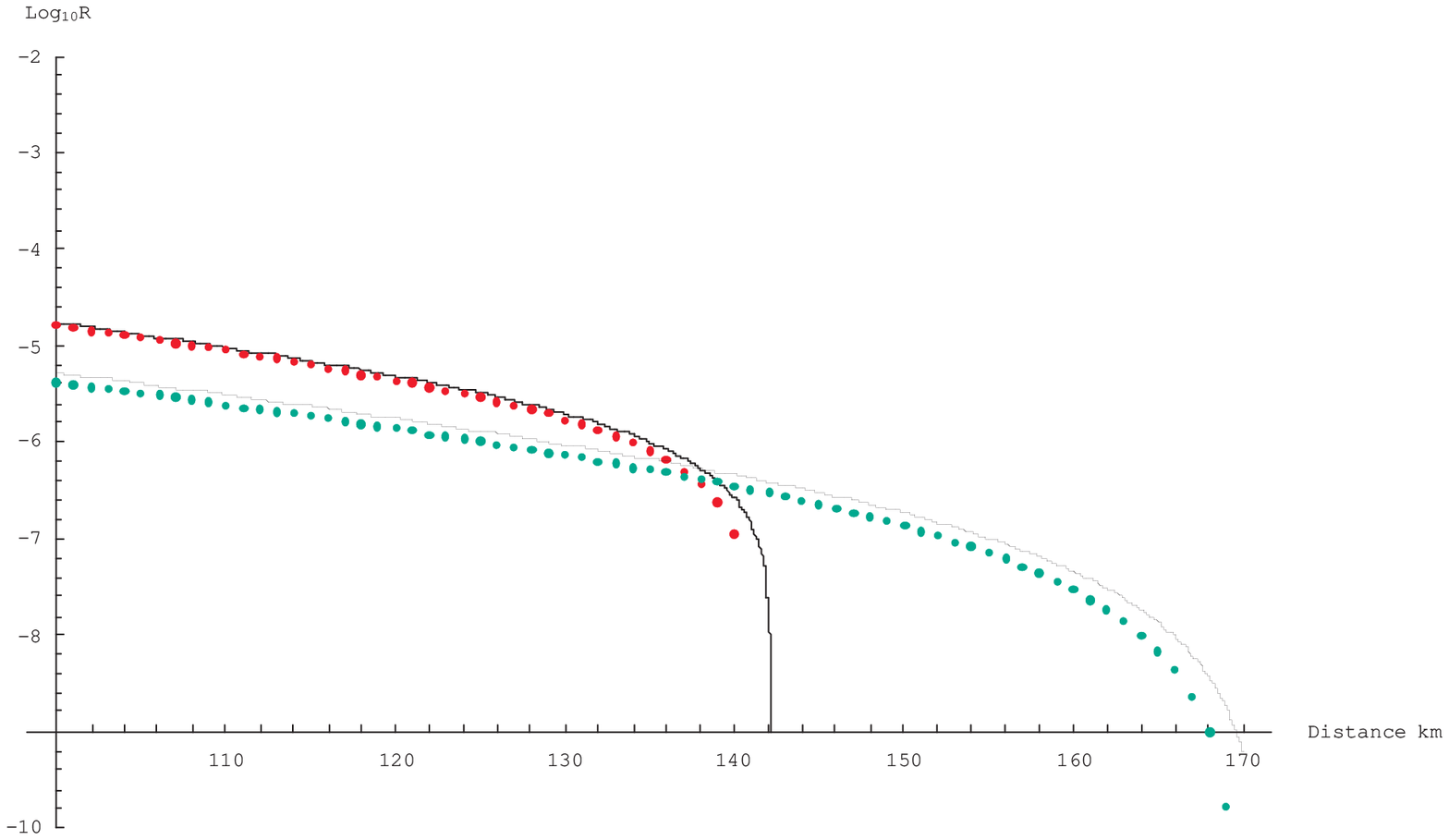}
\end{center}
\caption{Fig3. Key rate vs transmission distance. It shows the same objects as
in Fig. 2, but here we have set $\eta_{A}=0.6$.}%
\label{Fig3}%
\end{figure}

Fig. 2 shows different key generation rate comparing HSPS ($\eta_{A}=0.8$)
with WCS.

Fig. 3 shows the same objects as in Fig. 2 except with $\eta_{A}=0.6$.

From these simulations above, we can see that, our proposal can significantly
raise the transmission distance compared with that of coherent states even
with imperfect triggering detector ($\eta_{A}=0.6$). However, it has a lower
key generation rate. The reasons are as follows: On one hand, in HSPS the dark
count probability is so low by Alice's triggering system, there the
transmission distance is raised, on the other hand, an HSPS is basically a
thermal field which has a higher multi-photon probability than that in
Poissonian distribution with the same mean photon number, therefore the key
rate is decreased.

%To enlarge the final key generation rate, we can consider using a
%sub-Poissonian source instead of the thermal one. In fact, it can be realized
%by using a shutter at the exit of the source. (Which has already been
%mentioned in Ref. \cite{hori}.)

\section{Concluding remark.}

In summary, we have presented a practical decoy state method in quantum key
distribution with a heralded single photon source. By using 3 intensities, 0,
$\mu,~\mu^{\prime}$, we can estimate the lower bound of single-photon counts
and the upper bound of single-photon QBER rather tightly. Moreover, our
simulation results show that, the transmission distance of our 3-intensity
decoy-state QKD protocol with HSPS is larger than that of with weak coherence
states. Therefore, our 3-intensity decoy state method with HSPS seems to be a
promising candidate in practical implementation of quantum key distribution.

\begin{center}
\textbf{Acknowledgement}
\end{center}

Qin thanks Tomoyuki Horikiri (Univ. Tokyo) and Prof. Norbert L\"utkenhaus
(Univ. Waterloo) for useful discussions. This work is supported by National
Fundamental Research Program (2001CB309300), Tsinghua Bairen Program, the
Innovation Funds from Chinese Academy of Sciences, Program for New Century
Excellent Talents in University.

\end{document}